\documentclass[doublecol]{epl2} 
\usepackage[colorlinks=true,linkcolor=blue,urlcolor=blue,citecolor=blue]{hyperref}
\usepackage{amsmath}

\usepackage{mathtools}
\usepackage{mathrsfs}
\usepackage{csquotes}
\usepackage{braket}
\usepackage{amssymb}

\newcommand{\beq}{\begin{equation}}
\newcommand{\eeq}{\end{equation}}
\newcommand{\bea}{\begin{eqnarray}}
\newcommand{\eea}{\end{eqnarray}}

\title{\vspace*{-1cm}Quantum delocalization, structural order, and density response of the strongly coupled electron liquid}
\shorttitle{Quantum de-localization and structural order in the strongly coupled electron liquid} 

\author{Tobias Dornheim\inst{1,2,\#} \and Panagiotis Tolias\inst{3} \and Jan Vorberger\inst{2} \and Zhandos A. Moldabekov\inst{1,2}}
\shortauthor{T. Dornheim \etal}

\institute{                    
  \inst{1} Center for Advanced Systems Understanding (CASUS), D-02826 G\"orlitz, Germany\\
  \inst{2} Helmholtz-Zentrum Dresden-Rossendorf (HZDR), D-01328 Dresden, Germany\\
  \inst{3} Space and Plasma Physics, Royal Institute of Technology (KTH), Stockholm SE-100 44, Sweden\\
  \inst{\#} Electronic mail: t.dornheim@hzdr.de
}

\abstract{
We investigate the impact of electronic correlations and quantum delocalization onto the static structure factor and static density response function of the strongly coupled electron liquid. In contrast to a classical system, the density response of the electron liquid vanishes on small length scales due to quantum delocalization effects, which we rigorously quantify in terms of imaginary-time correlation functions and dynamic Matsubara response functions. This allows us to analyze the interplay of structural order and dynamic quantum effects as it manifests in the dynamic Matsubara local field correction. Finally, we identify an effective electronic attraction in the spin-offdiagonal static density response when the wavelength of the perturbation is commensurate with the average interparticle distance.
}

\begin{document}

\maketitle

\section{Introduction}

The uniform electron gas (UEG)~\cite{quantum_theory,loos,review} constitutes one of the most fundamental model systems in physics and quantum chemistry. Being often viewed as the archetypal system of interacting electrons, its properties are of prime importance for a broad variety of applications both at ambient conditions~\cite{Ceperley_Alder_PRL_1980,moroni2,cdop,vwn} and at extreme conditions~\cite{review,groth_prl,dornheim_prl,ksdt,Karasiev_status_2019,dornheim_ML} as they occur e.g.~in astrophysical objects and inertial fusion energy applications~\cite{drake2018high,fortov_review}. Indeed, the pressing need to understand the behaviour of electrons in this \emph{warm dense matter} (WDM) regime~\cite{Dornheim_review,wdm_book,new_POP} has facilitated a number of important developments in the field of fermionic quantum Monte Carlo (QMC) simulations at finite temperatures~\cite{Dornheim_POP_2017,review,Brown_PRL_2013,Malone_PRL_2016,Dornheim_NJP_2015,Joonho_JCP_2021,Hou_PRB_2022}.

Over the last years, particular interest has also emerged in the properties of strongly coupled electron liquids~\cite{Takada_PRB_2016,tanaka_hnc,dornheim_electron_liquid,dornheim_dynamic,Dornheim_Nature_2022,Dornheim_Force_2022,Tolias_JCP_2021,Tolias_JCP_2023,castello2021classical,Tolias_PRB_2024,koskelo2023shortrange}, i.e., the UEG at low densities $r_s\gtrsim10$ where $r_s=d/a_\textnormal{B}$ [with $d$ and $a_\textnormal{B}$ the Wigner-Seitz radius and Bohr radius]. While being hard to realize in contemporary experiments, this rich-in-physics regime is interesting in its own right. For example, it gives rise to a \emph{roton-type} feature in the dynamic structure factor~\cite{dornheim_dynamic,Takada_PRB_2016,koskelo2023shortrange,Dornheim_Nature_2022}, which has been interpreted in terms of the spontaneous alignment of pairs of electrons~\cite{Dornheim_Nature_2022,Dornheim_Force_2022} or, alternatively, as a consequence of excitonic attraction~\cite{koskelo2023shortrange}.
Having originally been reported for the strongly coupled UEG, the roton feature has subsequently been predicted to manifest in warm dense hydrogen~\cite{Hamann_PRR_2023} as it can be realized in hydrogen jets~\cite{Zastrau}, thereby opening up pathways for its experimental observation.
A further advantage of the electron liquid is that it allows one to focus on the nontrivial interplay of electronic correlations with quantum degeneracy effects, without the need to include other effects such as partial ionization and localization around ions~\cite{Dornheim_Science_2024}. This has facilitated a number of important developments in the field of dielectric theories~\cite{stls,stls_original,stls2,tanaka_hnc,Tolias_PRB_2024,Tolias_JCP_2021,Tolias_JCP_2023,arora}, which, in turn, inform practical applications such as the interpretation of X-ray Thomson scattering (XRTS) experiments~\cite{Dornheim_PRL_2020_ESA,Fortmann_PRE_2010,Zan_PRE_2021} and the development of the linear response time-dependent density functional theory for computational material science \cite{Runge_Gross_prl_1984, Byun_2020, Moldabekov_PRR_2023, Moldabekov_non_empirical_hybrid}.

In the present work, we carry out highly accurate \emph{ab initio} path integral Monte Carlo (PIMC) simulations of the spin unpolarized (paramagnetic) UEG at the electronic Fermi temperature $\Theta=k_\textnormal{B}T/E_\textnormal{F}=1$~\cite{quantum_theory,Ott2018} [where $E_\textnormal{F}$ is the Fermi energy] to study the impact of strong electronic correlations ($r_s=20,100,200$) onto the static structure factor $S(\mathbf{q})$ and the static linear density response function $\chi(\mathbf{q})$. Together with the utilization of the discrete imaginary Matsubara frequency space,
this gives us new insights into the interplay between quantum de-localization and electronic correlations. Moreover, we identify the second harmonic of the previously reported roton-type feature on both properties and discuss its impact on the (spin-resolved) density response, and its manifestation as a weak yet significant electronic attraction.

\begin{figure*}[t!]\centering\hspace*{0.5cm}
\includegraphics[width=0.4285\textwidth]{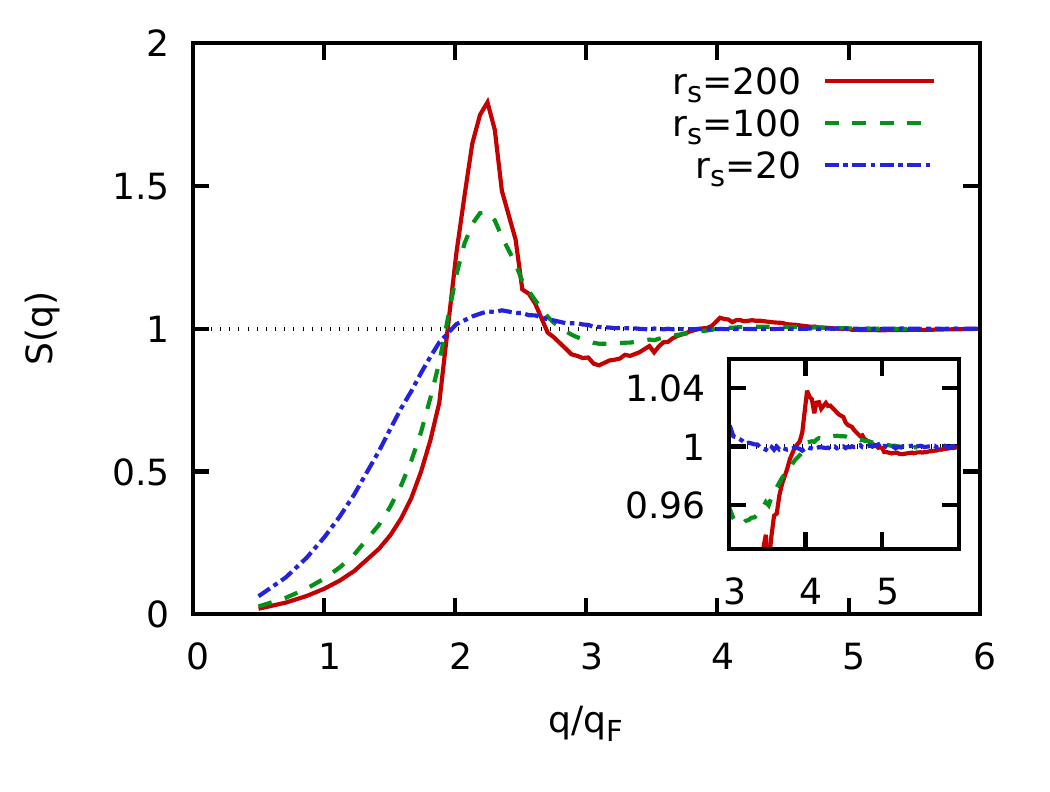}
\includegraphics[width=0.4285\textwidth]{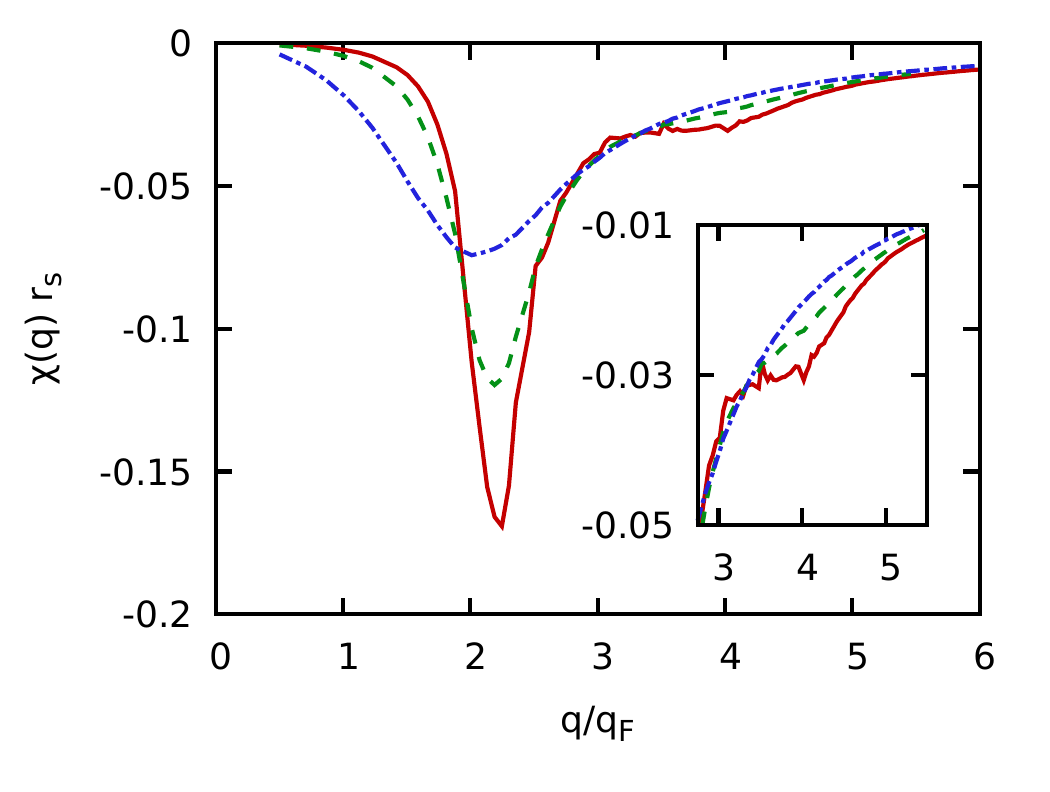}
\caption{\label{fig:compare} \emph{Ab initio} PIMC results for the static structure factor $S(\mathbf{q})$ [left] and the static linear density response function $\chi(\mathbf{q})$ [right] of the spin unpolarized electron liquid at the Fermi temperature $\Theta=1$ and $r_s=20,\,100,\,200$, with $N=66$ electrons.
}
\end{figure*}

\section{Theory}

Detailed introductions to the PIMC method have been presented in the literature~\cite{cep,boninsegni1} and need not be repeated here.
One of its key strengths is that it gives one access to the imaginary-time (density--density) correlation function (ITCF) $F(\mathbf{q},\tau)=\braket{\hat{n}(\mathbf{q},\tau)\hat{n}(-\mathbf{q},0)}$, with $\tau\in[0,\beta]$, $\beta=1/k_\textnormal{B}T$, and $t=-i\hbar\tau$ the imaginary time. An important property of the ITCF is its relation to the dynamic structure factor~\cite{Dornheim_review}
\begin{eqnarray}\label{eq:Laplace}
F(\mathbf{q},\tau) = \int_{-\infty}^\infty \textnormal{d}\omega\ S(\mathbf{q},\omega)\ e^{-\hbar\omega\tau}\ ,
\end{eqnarray}
which constitutes the basis for the \emph{analytic continuation}, i.e., the numerical inversion of Eq.~(\ref{eq:Laplace}) to obtain $S(\mathbf{q},\omega)$ based on PIMC results for the ITCF. The analytic continuation is an ill-posed problem in general~\cite{JARRELL1996133}, although particular solutions exist for the UEG~\cite{dornheim_dynamic,dynamic_folgepaper,Dornheim_PRE_2020} based on additional rigorous constraints. Interestingly, Eq.~(\ref{eq:Laplace}) constitutes the basis for the model-free interpretation of XRTS experiments with extreme states of matter~\cite{Dornheim_T_2022,Dornheim_T2_2022,dornheim2023xray}, as $F(\mathbf{q},\tau)$ contains, by definition, the same information as $S(\mathbf{q},\omega)$~\cite{Dornheim_insight_2022}, whereas the deconvolution with respect to the source and instrument function is substantially more stable in the Laplace domain.

The second ITCF relation that we utilize in the present work is the inverted version of the Fourier-Matsubara expansion derived by Tolias~\emph{et al.}~\cite{tolias2024fouriermatsubara}
\begin{eqnarray}\label{eq:MDR}
    \widetilde{\chi}(\mathbf{q},z_l) = -n\int_0^{\beta}\textnormal{d}\tau\ F(\mathbf{q},\tau)\ \textnormal{cos}\left(i z_l \tau\right)\ ,
\end{eqnarray}
with $n$ the number density and $\widetilde{\chi}(\mathbf{q},z_l)$ the dynamic density response at the discrete imaginary Matsubara frequencies $z_l=i 2\pi{l}/(\beta\hbar)$, with $l\in \mathbb{N}$. Note that the \enquote{tilde} symbol signifies dynamic quantities whose definition has been extended in the complex frequency domain by means of analytic continuation. Indeed, $\widetilde{\chi}(\mathbf{q},z_l)$ constitutes the central property in dielectric theories~\cite{stls}. Obviously, Eq.~(\ref{eq:MDR}) reverts to the usual imaginary-time version of the fluctuation--dissipation theorem~\cite{Dornheim_insight_2022} in the static limit of $l=0$,
\begin{eqnarray}\label{eq:static_chi}
    \chi(\mathbf{q}) = \widetilde{\chi}(\mathbf{q},0) = -n \int_0^\beta \textnormal{d}\tau\ F(\mathbf{q},\tau) \ .
\end{eqnarray}
From a theoretical perspective, it is convenient to express $\widetilde{\chi}(\mathbf{q},z_l)$ as~\cite{kugler1}
\begin{eqnarray}\label{eq:define_G}
    \widetilde{\chi}(\mathbf{q},z_l) = \frac{\widetilde{\chi}_0(\mathbf{q},z_l)}{1-4\pi{e}^2/q^2\left[1-\widetilde{G}(\mathbf{q},z_l)\right]\widetilde{\chi}_0(\mathbf{q},z_l)}\ ,
\end{eqnarray}
where $\widetilde{\chi}_0(\mathbf{q},z_l)$ is the dynamic Matsubara density response of a noninteracting Fermi gas. The complete wave-number and Matsubara-frequency resolved information about electronic correlations is thus contained in the dynamic local field correction $\widetilde{G}(\mathbf{q},z_l)$.

Finally, we note that the static linear density response function is related to the dynamic structure factor via its inverse moment sum-rule~\cite{Vitali_PRB_2010}
\begin{eqnarray}\label{eq:inverse_moment}
    \chi(\mathbf{q}) = - 2n \int_{-\infty}^\infty \textnormal{d}\omega\ \frac{S(\mathbf{q},\omega)}{\omega}\ .
\end{eqnarray}
In practice, Eqs.~(\ref{eq:Laplace}) and (\ref{eq:inverse_moment}) indicate a key difference between the static structure factor $S(\mathbf{q})=F(\mathbf{q},0)$ and the static density response $\chi(\mathbf{q})$: the former is completely insensitive to the peak position in $S(\mathbf{q},\omega)$, whereas the latter is highly sensitive and increases in magnitude when spectral weight is shifted to lower frequencies.

\section{Results}

All PIMC results have been obtained using the extended ensemble sampling scheme~\cite{Dornheim_PRB_nk_2021} as it has been implemented in the \texttt{ISHTAR} code~\cite{ISHTAR}; they are freely available in an online repository~\cite{repo}.

\begin{figure*}[t!]\centering\hspace*{0.5cm}
\includegraphics[width=0.4285\textwidth]{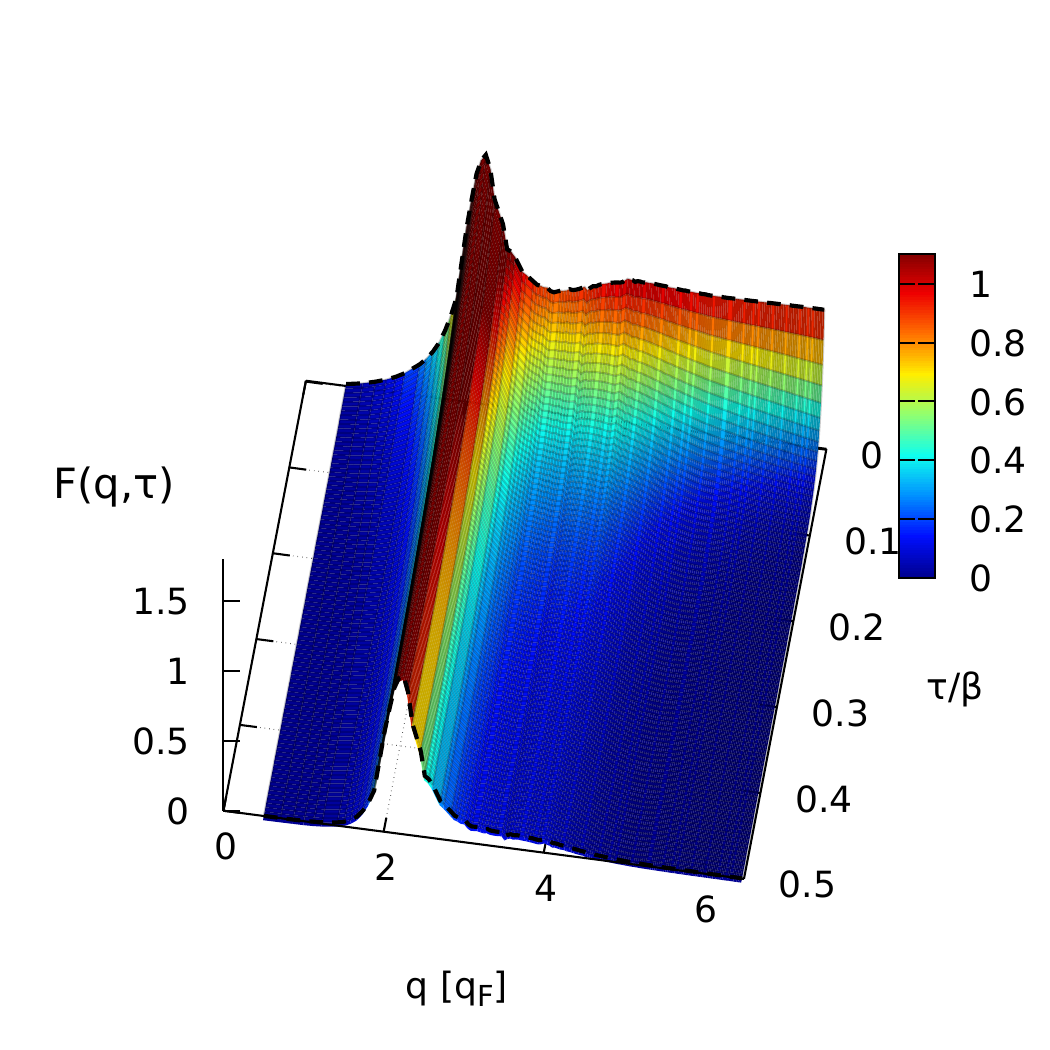}\hspace*{0.5cm}
\includegraphics[width=0.4285\textwidth]{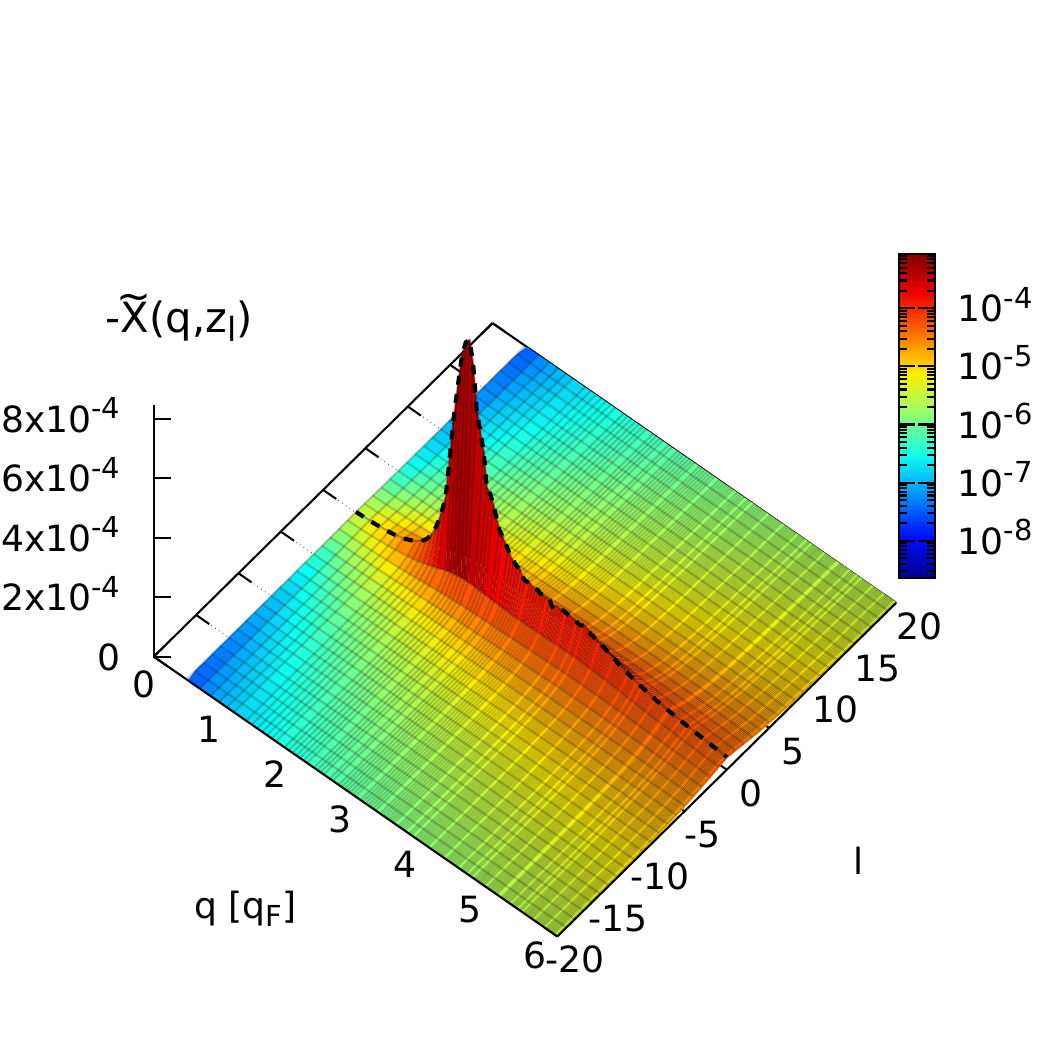}
\caption{\label{fig:imaginary} PIMC results for the ITCF $F(\mathbf{q},\tau)$ [left] and dynamic Matsubara density response function $\widetilde{\chi}(\mathbf{q},z_l)$ [right, cf.~Eq.~(\ref{eq:MDR})] of the spin unpolarized UEG at $r_s=200$ and $\Theta=1$. The dashed black lines in the left panel indicate the static structure factor $S(\mathbf{q})=F(\mathbf{q},0)$ and the thermal structure factor $F(\mathbf{q},\beta/2)$, respectively. The dashed black line in the right panel indicates the static density response function $\chi(\mathbf{q})$.}
\end{figure*} 

In the left panel of Fig.~\ref{fig:compare}, we show PIMC results for the static structure factor for $r_s=20$, $r_s=100$, and $r_s=200$ at $\Theta=1$. In the long-wavelength limit of $q\to0$, it holds $S(\mathbf{q})\sim q^2$ due to the perfect screening in the UEG~\cite{kugler_bounds}, whereas it holds $\lim_{q\to\infty}S(\mathbf{q})=1$ in the short-wavelength limit. The relevant information about electronic correlations and their impact on structural order thus manifests itself for intermediate $q$, and we find a pronounced peak around $q\approx2.1q_\textnormal{F}$ that increases with $r_s$. For $r_s=100$ and, in particular, for $r_s=200$, the first peak is followed by a minimum and a second maximum [see the inset], which is located around twice the position of the first peak.
Indeed, it is well known that $S(\mathbf{q})$ starts to exhibit increasing oscillations around $1$ with increasing coupling strength~\cite{Baus_Hansen_OCP,plasma2,Tolias_IEMHNC_2019}.

\begin{figure*}[t!]\centering
\includegraphics[width=0.39\textwidth]{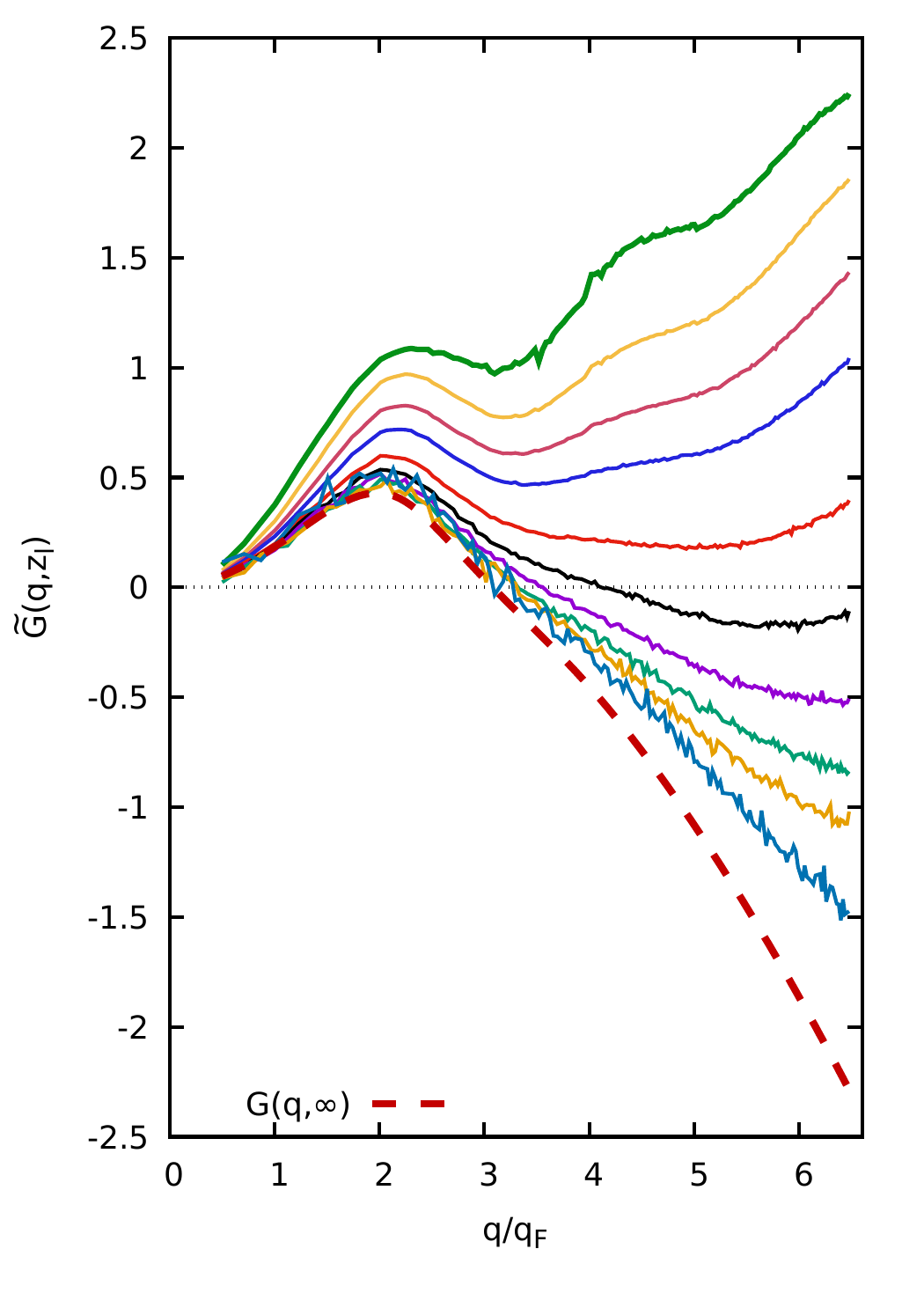}\includegraphics[width=0.39\textwidth]{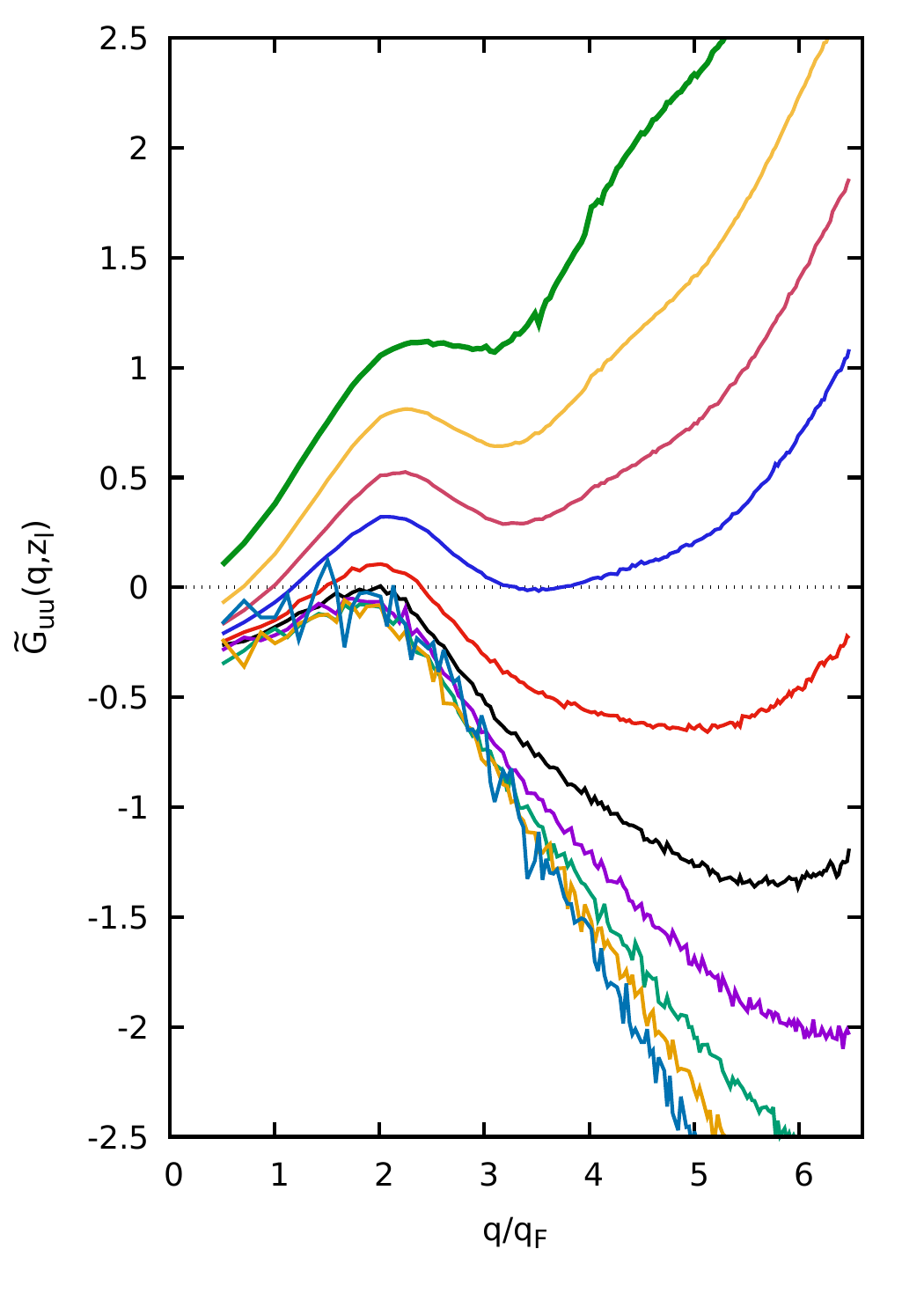}\\
\includegraphics[width=0.39\textwidth]{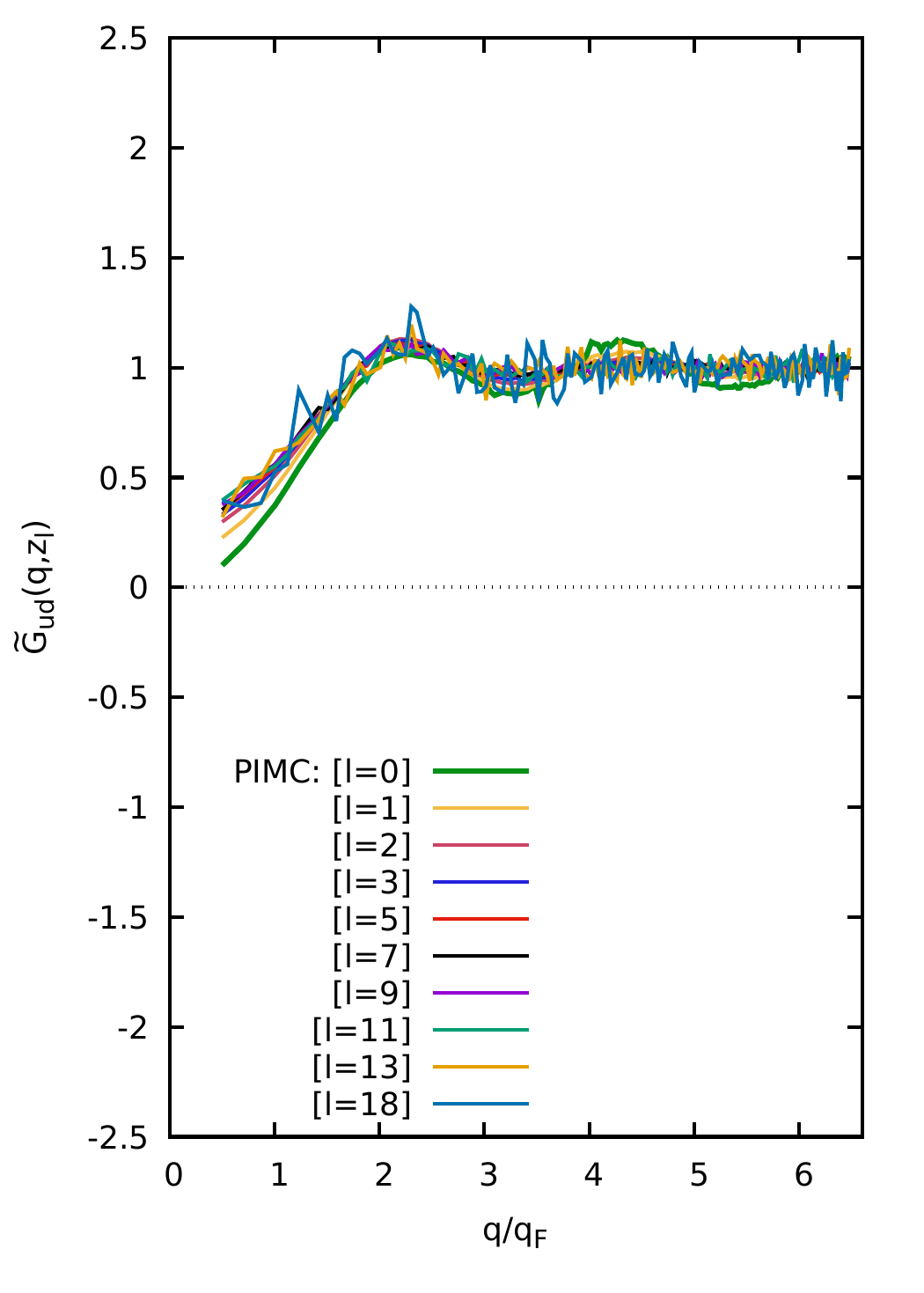}
\includegraphics[width=0.39\textwidth]{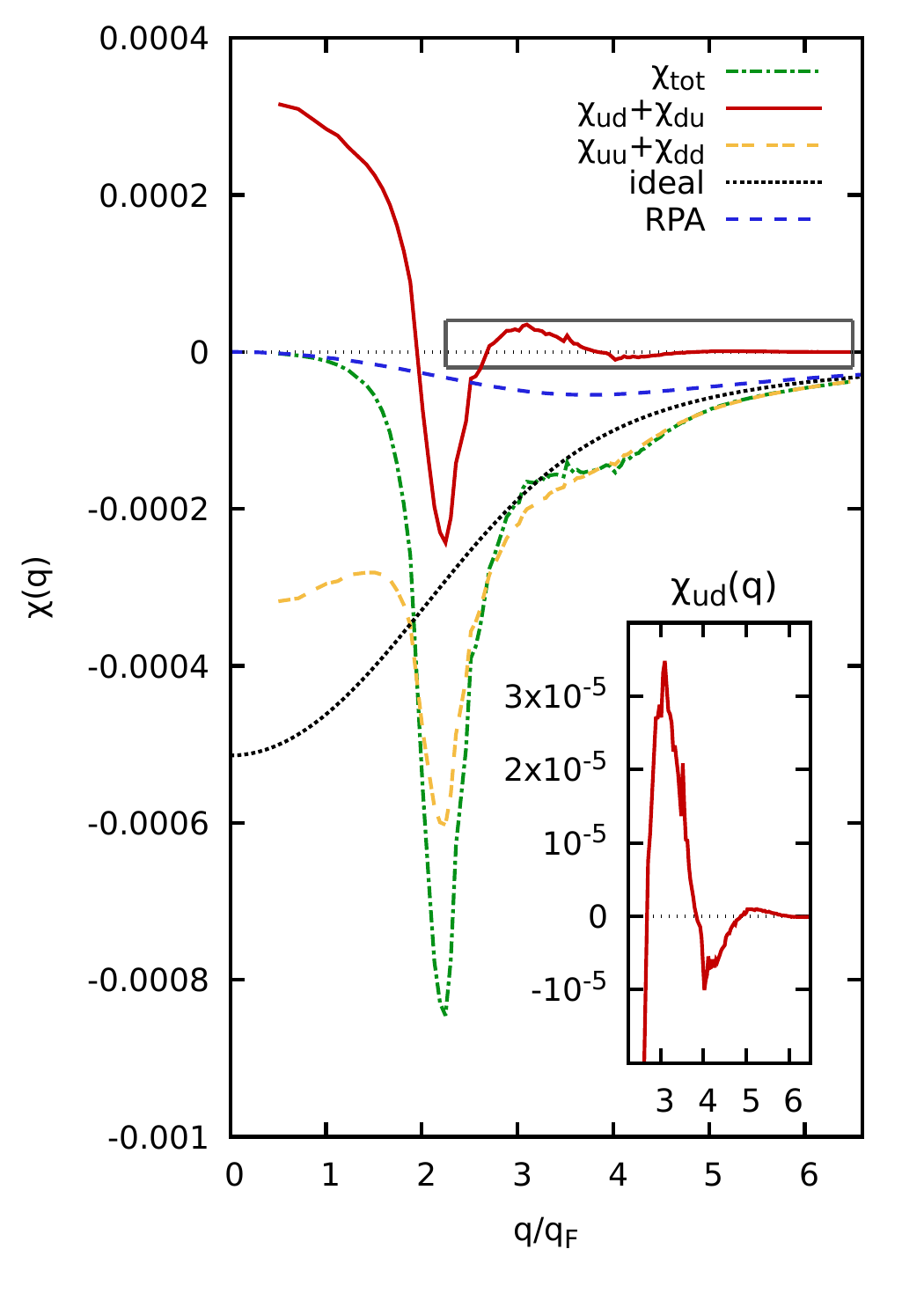}
\caption{\label{fig:PIMC} Top and bottom left: Wavenumber dependence of the total $\widetilde{G}(\mathbf{q},z_l)$, spin-diagonal $\widetilde{G}_{\mathrm{uu}}(\mathbf{q},z_l)$, and spin-offdiagonal $\widetilde{G}_{\mathrm{ud}}(\mathbf{q},z_l)$ dynamic Matsubara local field corrections of the spin unpolarized UEG at $r_s=200$ and $\Theta=1$. Bottom right: various static density response functions.
}
\end{figure*}

In the right panel of Fig.~\ref{fig:compare}, we show the corresponding static density response function $\chi(\mathbf{q})$ at the same parameters. For a classical system, $\chi(\mathbf{q})$ would exhibit exactly the same behaviour as $S(\mathbf{q})$, since they are related by~\cite{kugler_classical}
\begin{eqnarray}\label{eq:classic}
    \chi^\textnormal{cl}(\mathbf{q}) = - n\beta S^\textnormal{cl}(\mathbf{q})\ .
\end{eqnarray}
For a quantum mechanical system, on the other hand, the product $\beta{S}^\textnormal{cl}(\mathbf{q})$ is replaced by an integral over the ITCF, which decays with $\tau$ [cf.~Fig.~\ref{fig:imaginary}]. A rigorous result of general linear response theory states that, for any Hermitian operator, the classical limit of the static response function serves as an upper bound (in absolute values) of the quantum static response function~\cite{quantum_theory}. This is formally derived on the basis of the convenient Lehmann representation, but it can also be derived on this basis of the imaginary-time version of the fluctuation--dissipation theorem, see Eq.~(\ref{eq:static_chi}), after noting that $0\leq{F}(\mathbf{q},\tau)\leq{F}(\mathbf{q},0)=S(\mathbf{q})$ courtesy of the imaginary-time diffusion~\cite{Dornheim_PTR_2022}. Thus, the response of a quantum system is systematically reduced compared to the classical case due to quantum delocalization. In particular, a quantum system will stop responding to an external harmonic perturbation all together when the response wavelength $\lambda_q=2\pi/q$ becomes much smaller than its thermal wavelength $\lambda_\beta=\sqrt{2\pi\hbar^2\beta/m}$. In practice, we find that the PIMC results for $\chi(\mathbf{q})$ exhibit less structure compared to $S(\mathbf{q})$, although there does appear a saddle point around $q=4q_\textnormal{F}$ for $r_s=200$.

In Fig.~\ref{fig:imaginary}, we provide the full imaginary-time structure of the density--density correlations for $r_s=200$ and $\Theta=1$.
The left panel shows the ITCF, which is symmetric around $\tau=\beta/2$, i.e., $F(\mathbf{q},\tau)=F(\mathbf{q},\beta-\tau)$; this is equivalent to the well-known detailed balance relation of the dynamic structure factor $S(\mathbf{q},-\omega)=e^{-\beta\hbar\omega}S(\mathbf{q},\omega)$~\cite{quantum_theory} and is of practical importance for the interpretation of XRTS experiments~\cite{Dornheim_T_2022,Dornheim_T2_2022}. The dashed black line corresponds to the static structure factor $S(\mathbf{q})=F(\mathbf{q},0)$ shown above. The ITCF then decays for $0\leq\tau\leq\beta/2$ due to thermal quantum de-localization as it has been explained in detail in Ref.~\cite{Dornheim_PTR_2022}. The minimum at $F(\mathbf{q},\beta/2)$ has been coined as the \emph{thermal structure factor} in the literature~\cite{Dornheim_insight_2022} and also exhibits an interesting dependence on $q$. First, it attains a minimum at the same position as $S(\mathbf{q})$, for the same reasons. Second, it exhibits a plateau around $q=4q_\textnormal{F}$, which is responsible for the corresponding saddle point in $\chi(\mathbf{q})$.

In the right panel of Fig.~\ref{fig:imaginary}, we show the dynamic Matsubara density response function in the $q$-$l$-plane, with $l$ being the index (order) of the discrete Matsubara frequencies. Again, the dashed black line shows the static limit of $\widetilde{\chi}(\mathbf{q},0)=\chi(\mathbf{q})$, and $\widetilde{\chi}(\mathbf{q},l)$ exhibits a rapid decay with $l$. This decay becomes less pronounced with increasing $q$, as quantum de-localization effects become more important on smaller length scales $\lambda_q$. We note that $S(\mathbf{q})$ and $\widetilde{\chi}(\mathbf{q},z_l)$ are related by~\cite{stls}
\begin{eqnarray}\label{eq:sum}
    S(\mathbf{q}) = - \frac{1}{n\beta}\sum_{l=-\infty}^\infty \widetilde{\chi}(\mathbf{q},z_l)\ .
\end{eqnarray}
While $\chi(\mathbf{q})$ vanishes for large $q$ as de-localized particles stop reacting to an external perturbation when $\lambda_\beta \gg \lambda_q$, the total spectral weight of the dynamic Matsubara density response is still conserved in Eq.~(\ref{eq:sum}) as $S(\mathbf{q})$ attains unity in this limit.

To resolve the impact of dynamic correlations, we show the wavenumber dependence of the dynamic Matsubara local field correction for various orders $l$ in the top left panel of Fig.~\ref{fig:PIMC}. These results have been obtained by inverting Eq.~(\ref{eq:define_G}) based on our quasi-exact PIMC results for $\widetilde{\chi}(\mathbf{q},z_l)$. In the static limit (dark green), the behaviour of $\widetilde{G}(\mathbf{q},0)=G(\mathbf{q})$ for small $q$ is governed by the well-known compressibility sum-rule~\cite{Dornheim_PRB_ESA_2021,quantum_theory} that presets a parabolic increase. This is followed by a maximum around $q=2q_\textnormal{F}$, and a pronounced saddle point at $q=4q_\textnormal{F}$. These features become less pronounced for larger $l$, but never vanish entirely; they are even noticeable in the high-frequency limit $\widetilde{G}(\mathbf{q},\infty)$~\cite{quantum_theory,kugler1} that is included as the dashed red line. It is thus clear that the impact of correlation effects on $S(\mathbf{q})$ and $\chi(\mathbf{q})$ is not purely static in origin, but also governs the dynamic behaviour of the system. This is consistent with previous dielectric theory investigations that have reported the importance of a dynamic local field correction in the strongly coupled electron liquid regime~\cite{Tolias_JCP_2023}. For every finite $l$, $\widetilde{G}(\mathbf{q},z_l)$ eventually increases parabolically with $q$, see Ref.~\cite{Hou_PRB_2022}. Conversely, $\widetilde{G}(\mathbf{q},z_l)$ converges towards its high-frequency limit (dashed red) with $l$ from above for every fixed wavenumber, see also the recent discussion in Ref.~\cite{dornheim2024MatsubaraResponse}.

\begin{figure}[t!]\centering
\includegraphics[width=0.4285\textwidth]{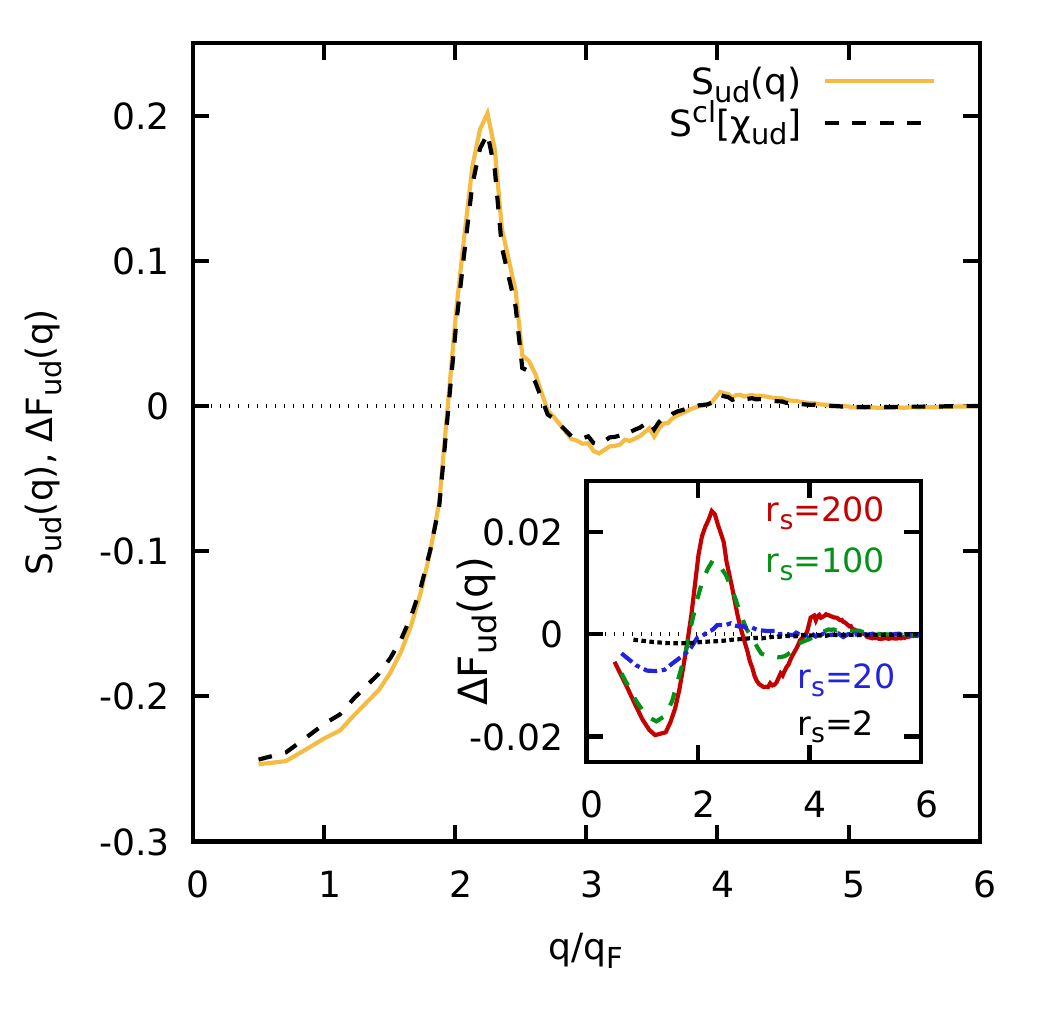}
\caption{\label{fig:delta} Main: spin-offdiagonal static structure factor $S_{\mathrm{ud}}(\mathbf{q})$ [solid yellow] and classical relation Eq.~(\ref{eq:classic}) [dashed black] for the spin unpolarized UEG at $r_s=200$ and $\Theta=1$. Inset: spin-offdiagonal imaginary-time decay $\Delta F_{\mathrm{ud}}(q)=F_{\mathrm{ud}}(\mathbf{q},0)-F_{\mathrm{ud}}(\mathbf{q},\beta/2)$ at $r_s=200$ [solid red], $r_s=100$ [dashed green], $r_s=20$ [dash-dotted blue], and $r_s=2$ [dotted black, taken from Ref.~\cite{Dornheim_PRR_2022}]. 
}
\end{figure}

To further analyze the observed trends, we show the spin-diagonal and spin-offdiagonal dynamic Matsubara local field corrections in the top right and bottom left panels of Fig.~\ref{fig:PIMC}. Note the exact connection $\widetilde{G}\equiv(\widetilde{G}_{\mathrm{ud}}+\widetilde{G}_{\mathrm{uu}})/2$ that is valid in the spin unpolarized UEG case, as a consequence of $\widetilde{G}_{\mathrm{ud}}=\widetilde{G}_{\mathrm{du}}$ (reciprocity), $\widetilde{G}_{\mathrm{dd}}=\widetilde{G}_{\mathrm{uu}}$ (equal spin-up and spin-down densities) ~\cite{quantum_theory,Dornheim_PRR_2022}. It is evident that $\widetilde{G}_{\mathrm{ud}}(\mathbf{q},z_l)$ qualitatively resembles the full $\widetilde{G}(\mathbf{q},z_l)$ and exhibits similar short wavelength asymptotics. In stark contrast, $\widetilde{G}_{\mathrm{ud}}(\mathbf{q},z_l)$ exhibits a substantially less pronounced, though still significant, dependence on $l$. In the static limit, $\widetilde{G}_{\mathrm{ud}}(\mathbf{q},0)$ exhibits oscillations that i) exhibit maxima at the same positions as $S(\mathbf{q})$ [and $S_{\mathrm{ud}}(\mathbf{q})$, cf.~Fig.~\ref{fig:delta} below] and ii) only weakly decay with $q$. We point out that the weak dependence of $\widetilde{G}_{\mathrm{ud}}(\mathbf{q},z_l)$ on $l$ implies that static dielectric theories might be sufficient to describe spin-offdiagonal density correlations even in the strongly coupled electron liquid.

In the bottom right panel of Fig.~\ref{fig:PIMC}, we show different components of the full static density response, which gives us additional insights into the implication and origin of the observed effect. The dashed blue curve shows the random phase approximation that is computed by setting $\widetilde{G}(\mathbf{q},z_l)\equiv 0$ in Eq.~(\ref{eq:define_G}); it agrees only in the limits of small and large $q$ with the full PIMC curve (dash-dotted green).
Interestingly, $\chi_0(\mathbf{q})$ [dotted black] is in better agreement with PIMC for $q\gtrsim3q_\textnormal{F}$, although it, too, misses the correlation features by definition.
The solid red and double-dashed yellow curves show the spin-offdiagonal and spin-diagonal static density response functions~\cite{Dornheim_PRR_2022}. Overall, $\chi_{\mathrm{uu}}(\mathbf{q})=\chi_{\mathrm{dd}}(\mathbf{q})$ (spin unpolarized UEG) qualitatively follows the full response function, except in the limit of small $q$; here the full $\chi(\mathbf{q})$ exhibits the familiar perfect screening~\cite{kugler_bounds} as $\chi_{\mathrm{ud}}(\mathbf{q})+\chi_{\mathrm{du}}(\mathbf{q)}$ and $\chi_{\mathrm{dd}}(\mathbf{q})+\chi_{\mathrm{uu}}(\mathbf{q})$ exactly cancel. However, the spin-resolved components do not vanish individually~\cite{Dornheim_PRR_2022}. The spin-offdiagonal component, $\chi_{\mathrm{ud}}(\mathbf{q})=\chi_{\mathrm{du}}(\mathbf{q})$ (reciprocity), is even more interesting and deserves special attention. Assume an external harmonic perturbation that only acts on the spin-up electrons. For small $q$, the spin-down electrons will move in the opposite direction as the spin-up electrons, i.e., towards the maxima of the external perturbation; this leads to the perfect screening of the total density response. Around $q=2q_\textnormal{F}$, the wavelength of the external perturbation is commensurate to the average inter-particle distance, $\lambda_q\approx1.6r_s$. It is thus advantageous for the unperturbed spin-down electrons to align themselves with the external perturbation to minimize the total free energy of the system. This is purely an exchange--correlation effect that is not even qualitatively included in the RPA~\cite{Dornheim_Nature_2022,Dornheim_Force_2022}.
The same trends repeat with increasing $q$, although with a substantially reduced magnitude; we find a significant negative minimum in $\chi_{\mathrm{ud}}(\mathbf{q})$ for $q=4q_\textnormal{F}$, and a very small residual minimum around $q=6q_\textnormal{F}$, see the inset.

Finally, we show the spin-offdiagonal static structure factor $S_{\mathrm{ud}}(\mathbf{q})$ in Fig.~\ref{fig:delta}. It strongly resembles $\chi_{\mathrm{ud}}(\mathbf{q})$ and the classical relation [Eq.~(\ref{eq:classic}), dashed black line] is quite accurate; this is expected as quantum delocalization is known to be less important for inter-species correlation functions~\cite{Dornheim_PRR_2022}. Nevertheless, we do find a small yet significant dependence of $F_{\mathrm{ud}}(\mathbf{q},\tau)$ on $\tau$, see the inset showing the difference $\Delta F_{\mathrm{ud}}(\mathbf{q}) = F_{\mathrm{ud}}(\mathbf{q},0)-F_{\mathrm{ud}}(\mathbf{q},\beta/2)$ over a broad range of $r_s$-values. Evidently, $\Delta F_{\mathrm{ud}}(\mathbf{q})$ strongly increases with $r_s$ in absolute terms, which is a consequence of the overall increased magnitude of $S_{\mathrm{ud}}(\mathbf{q})$ and $F_{\mathrm{ud}}(\mathbf{q},\tau)$.

\section{Discussion}

We have presented highly accurate \emph{ab initio} PIMC results for the static structure factor, static density response, imaginary-time correlation function, dynamic Matsubara density response $\widetilde{\chi}(\mathbf{q},z_l)$, and dynamic Matsubara local field correction $\widetilde{G}(\mathbf{q},z_l)$ of the strongly coupled electron liquid. In contrast to classical systems, structural-order based oscillations in $S(\mathbf{q})$ do not directly translate to $\chi(\mathbf{q})$, but vanish with increasing $q$ due to quantum de-localization effects. At the same time,  the structural order does impose an imprint onto the dynamic Matsubara local field correction $\widetilde{G}(\mathbf{q},z_l)$ [and its spin-resolved components $\widetilde{G}_{\mathrm{uu}}(\mathbf{q},z_l)$ and $\widetilde{G}_{\mathrm{ud}}(\mathbf{q},z_l)$] that even persists, although weakly, in the limit of infinite Matsubara order. Moreover, we have found higher-order minima in the spin-offdiagonal static density response that indicate an effective electronic attraction. Finally, we have analyzed the impact of quantum delocalization onto the spin-offdiagonal ITCF, which is substantially larger in absolute terms than in the previously investigated warm-dense matter regime~\cite{Dornheim_PRR_2022} due to the overall higher degree of correlation in the electron liquid.

In addition to being interesting in their own right, we expect our new results to be valuable for the further development of dielectric theories, and for the development of new tools for warm dense matter research. All results are freely available online~\cite{repo} and can be used to benchmark models and approximations. Moreover, we note that the present study can be complemented by an analytic continuation to obtain $S(\mathbf{q},\omega)$, which might reveal a second roton around the second harmonic of the original feature. Finally, we propose to study the nonlinear response of the strongly coupled electron liquid, which is connected to higher-order (e.g.~three-body for the quadratic response) correlation functions~\cite{Dornheim_JCP_ITCF_2021,Dornheim_JPSJ_2021}.



\acknowledgments
\section{Acknowledgements}
This work was partially supported by the Center for Advanced Systems Understanding (CASUS), financed by Germany’s Federal Ministry of Education and Research (BMBF) and the Saxon state government out of the State budget approved by the Saxon State Parliament. 
This work has received funding from the European Research Council (ERC) under the European Union’s Horizon 2022 research and innovation programme
(Grant agreement No. 101076233, "PREXTREME"). Views and opinions expressed are however those of the authors only and do not necessarily reflect those of the European Union or the European Research Council Executive Agency. Neither the European Union nor the granting authority can be held responsible for them. Computations were performed at the Norddeutscher Verbund f\"ur Hoch- und H\"ochstleistungsrechnen (HLRN) under grant mvp00024.

\bibliographystyle{iopart-num}
\bibliography{bibliography2.bib}

\providecommand{\newblock}{}
\begin{thebibliography}{10}
\expandafter\ifx\csname url\endcsname\relax
  \def\url#1{{\tt #1}}\fi
\expandafter\ifx\csname urlprefix\endcsname\relax\def\urlprefix{URL }\fi
\providecommand{\eprint}[2][]{\url{#2}}

\bibitem{quantum_theory}
Giuliani G and Vignale G 2008 {\em Quantum Theory of the Electron Liquid\/} (Cambridge: Cambridge University Press)

\bibitem{loos}
Loos P~F and Gill P~M~W 2016 {\em Comput. Mol. Sci\/} {\bf 6} 410--429

\bibitem{review}
Dornheim T, Groth S and Bonitz M 2018 {\em Phys. Rep.\/} {\bf 744} 1--86

\bibitem{Ceperley_Alder_PRL_1980}
Ceperley D~M and Alder B~J 1980 {\em Phys. Rev. Lett.\/} {\bf 45}(7) 566--569

\bibitem{moroni2}
Moroni S, Ceperley D~M and Senatore G 1995 {\em Phys. Rev. Lett\/} {\bf 75} 689

\bibitem{cdop}
Corradini M, Sole R~D, Onida G and Palummo M 1998 {\em Phys. Rev. B\/} {\bf 57} 14569

\bibitem{vwn}
Vosko S~H, Wilk L and Nusair M 1980 {\em Can. J. Phys.\/} {\bf 58} 1200--1211

\bibitem{groth_prl}
Groth S, Dornheim T, Sjostrom T, Malone F~D, Foulkes W~M~C and Bonitz M 2017 {\em Phys. Rev. Lett.\/} {\bf 119} 135001

\bibitem{dornheim_prl}
Dornheim T, Groth S, Sjostrom T, Malone F~D, Foulkes W~M~C and Bonitz M 2016 {\em Phys. Rev. Lett.\/} {\bf 117} 156403

\bibitem{ksdt}
Karasiev V~V, Sjostrom T, Dufty J and Trickey S~B 2014 {\em Phys. Rev. Lett.\/} {\bf 112}(7) 076403

\bibitem{Karasiev_status_2019}
Karasiev V~V, Trickey S~B and Dufty J~W 2019 {\em Phys. Rev. B\/} {\bf 99}(19) 195134

\bibitem{dornheim_ML}
Dornheim T, Vorberger J, Groth S, Hoffmann N, Moldabekov Z and Bonitz M 2019 {\em J. Chem. Phys\/} {\bf 151} 194104

\bibitem{drake2018high}
Drake R 2018 {\em High-Energy-Density Physics: Foundation of Inertial Fusion and Experimental Astrophysics\/} Graduate Texts in Physics (Springer International Publishing)

\bibitem{fortov_review}
Fortov V~E 2009 {\em Phys.-Usp\/} {\bf 52} 615--647

\bibitem{Dornheim_review}
Dornheim T, Moldabekov Z~A, Ramakrishna K, Tolias P, Baczewski A~D, Kraus D, Preston T~R, Chapman D~A, Böhme M~P, Döppner T, Graziani F, Bonitz M, Cangi A and Vorberger J 2023 {\em Phys. Plasmas\/} {\bf 30} 032705

\bibitem{wdm_book}
Graziani F, Desjarlais M~P, Redmer R and Trickey S~B (eds) 2014 {\em Frontiers and Challenges in Warm Dense Matter\/} (International Publishing: Springer)

\bibitem{new_POP}
Bonitz M, Dornheim T, Moldabekov Z~A, Zhang S, Hamann P, Kählert H, Filinov A, Ramakrishna K and Vorberger J 2020 {\em Phys. Plasmas\/} {\bf 27} 042710

\bibitem{Dornheim_POP_2017}
Dornheim T, Groth S, Malone F~D, Schoof T, Sjostrom T, Foulkes W~M~C and Bonitz M 2017 {\em Phys. Plasmas\/} {\bf 24} 056303

\bibitem{Brown_PRL_2013}
Brown E~W, Clark B~K, DuBois J~L and Ceperley D~M 2013 {\em Phys. Rev. Lett.\/} {\bf 110}(14) 146405

\bibitem{Malone_PRL_2016}
Malone F~D, Blunt N~S, Brown E~W, Lee D~K~K, Spencer J~S, Foulkes W~M~C and Shepherd J~J 2016 {\em Phys. Rev. Lett.\/} {\bf 117}(11) 115701

\bibitem{Dornheim_NJP_2015}
Dornheim T, Groth S, Filinov A and Bonitz M 2015 {\em New J. Phys.\/} {\bf 17} 073017

\bibitem{Joonho_JCP_2021}
Lee J, Morales M~A and Malone F~D 2021 {\em J. Chem. Phys.\/} {\bf 154} 064109

\bibitem{Hou_PRB_2022}
Hou P~C, Wang B~Z, Haule K, Deng Y and Chen K 2022 {\em Phys. Rev. B\/} {\bf 106}(8) L081126

\bibitem{Takada_PRB_2016}
Takada Y 2016 {\em Phys. Rev. B\/} {\bf 94}(24) 245106

\bibitem{tanaka_hnc}
Tanaka S 2016 {\em J. Chem. Phys.\/} {\bf 145} 214104

\bibitem{dornheim_electron_liquid}
Dornheim T, Sjostrom T, Tanaka S and Vorberger J 2020 {\em Phys. Rev. B\/} {\bf 101}(4) 045129

\bibitem{dornheim_dynamic}
Dornheim T, Groth S, Vorberger J and Bonitz M 2018 {\em Phys. Rev. Lett.\/} {\bf 121} 255001

\bibitem{Dornheim_Nature_2022}
Dornheim T, Moldabekov Z, Vorberger J, K{\"a}hlert H and Bonitz M 2022 {\em Commun. Phys.\/} {\bf 5} 304 ISSN 2399-3650

\bibitem{Dornheim_Force_2022}
Dornheim T, Tolias P, Moldabekov Z~A, Cangi A and Vorberger J 2022 {\em J. Chem. Phys.\/} {\bf 156} 244113

\bibitem{Tolias_JCP_2021}
Tolias P, Lucco~Castello F and Dornheim T 2021 {\em J. Chem. Phys.\/} {\bf 155} 134115

\bibitem{Tolias_JCP_2023}
Tolias P, Lucco~Castello F and Dornheim T 2023 {\em J. Chem. Phys.\/} {\bf 158} 141102

\bibitem{castello2021classical}
Lucco~Castello F, Tolias P and Dornheim T 2022 {\em Europhysics Letters\/} {\bf 138} 44003

\bibitem{Tolias_PRB_2024}
Tolias P, Lucco~Castello F, Kalkavouras F and Dornheim T 2024 {\em Phys. Rev. B\/} {\bf 109}(12) 125134

\bibitem{koskelo2023shortrange}
Koskelo J, Reining L and Gatti M 2023 Short-range excitonic phenomena in low-density metals (\textit{Preprint} \eprint{2301.00474})

\bibitem{Hamann_PRR_2023}
Hamann P, Kordts L, Filinov A, Bonitz M, Dornheim T and Vorberger J 2023 {\em Phys. Rev. Res.\/} {\bf 5}(3) 033039

\bibitem{Zastrau}
Zastrau U, Sperling P, Harmand M, Becker A, Bornath T, Bredow R, Dziarzhytski S, Fennel T, Fletcher L~B, F{\"o}rster E, G{\"o}de S, Gregori G, Hilbert V, Hochhaus D, Holst B, Laarmann T, Lee H~J, Ma T, Mithen J~P, Mitzner R, Murphy C~D, Nakatsutsumi M, Neumayer P, Przystawik A, Roling S, Schulz M, Siemer B, Skruszewicz S, Tiggesb{\"a}umker J, Toleikis S, Tschentscher T, White T, W{\"o}stmann M, Zacharias H, D{\"o}ppner T, Glenzer S~H and Redmer R 2014 {\em Phys. Rev. Lett\/} {\bf 112} 105002

\bibitem{Dornheim_Science_2024}
Dornheim T, Döppner T, Tolias P, Böhme M, Fletcher L, Gawne T, Graziani F, Kraus D, MacDonald M, Moldabekov Z, Schwalbe S, Gericke D and Vorberger J 2024 Unraveling electronic correlations in warm dense quantum plasmas (\textit{Preprint} \eprint{2402.19113})

\bibitem{stls}
Tanaka S and Ichimaru S 1986 {\em J. Phys. Soc. Jpn\/} {\bf 55} 2278--2289

\bibitem{stls_original}
Singwi K~S, Tosi M~P, Land R~H and Sj\"olander A 1968 {\em Phys. Rev\/} {\bf 176} 589

\bibitem{stls2}
Sjostrom T and Dufty J 2013 {\em Phys. Rev. B\/} {\bf 88} 115123

\bibitem{arora}
Arora P, Kumar K and Moudgil R~K 2017 {\em Eur. Phys. J. B\/} {\bf 90} 76

\bibitem{Dornheim_PRL_2020_ESA}
Dornheim T, Cangi A, Ramakrishna K, B\"ohme M, Tanaka S and Vorberger J 2020 {\em Phys. Rev. Lett.\/} {\bf 125}(23) 235001

\bibitem{Fortmann_PRE_2010}
Fortmann C, Wierling A and R\"opke G 2010 {\em Phys. Rev. E\/} {\bf 81}(2) 026405

\bibitem{Zan_PRE_2021}
Zan X, Lin C, Hou Y and Yuan J 2021 {\em Phys. Rev. E\/} {\bf 104}(2) 025203

\bibitem{Runge_Gross_prl_1984}
Runge E and Gross E~K~U 1984 {\em Phys. Rev. Lett.\/} {\bf 52}(12) 997--1000 \urlprefix\url{https://link.aps.org/doi/10.1103/PhysRevLett.52.997}

\bibitem{Byun_2020}
Byun Y~M, Sun J and Ullrich C~A 2020 {\em Electronic Structure\/} {\bf 2} 023002 \urlprefix\url{https://dx.doi.org/10.1088/2516-1075/ab7b12}

\bibitem{Moldabekov_PRR_2023}
Moldabekov Z~A, Pavanello M, B\"ohme M~P, Vorberger J and Dornheim T 2023 {\em Phys. Rev. Res.\/} {\bf 5}(2) 023089 \urlprefix\url{https://link.aps.org/doi/10.1103/PhysRevResearch.5.023089}

\bibitem{Moldabekov_non_empirical_hybrid}
Moldabekov Z~A, Lokamani M, Vorberger J, Cangi A and Dornheim T 2023 {\em The Journal of Physical Chemistry Letters\/} {\bf 14} 1326--1333 \urlprefix\url{https://doi.org/10.1021/acs.jpclett.2c03670}

\bibitem{Ott2018}
Ott T, Thomsen H, Abraham J~W, Dornheim T and Bonitz M 2018 {\em Eur. Phys. J. D\/} {\bf 72} 84

\bibitem{cep}
Ceperley D~M 1995 {\em Rev. Mod. Phys\/} {\bf 67} 279

\bibitem{boninsegni1}
Boninsegni M, Prokofev N~V and Svistunov B~V 2006 {\em Phys. Rev. E\/} {\bf 74} 036701

\bibitem{JARRELL1996133}
Jarrell M and Gubernatis J 1996 {\em Phys. Rep.\/} {\bf 269} 133--195

\bibitem{dynamic_folgepaper}
Groth S, Dornheim T and Vorberger J 2019 {\em Phys. Rev. B\/} {\bf 99} 235122

\bibitem{Dornheim_PRE_2020}
Dornheim T and Vorberger J 2020 {\em Phys. Rev. E\/} {\bf 102}(6) 063301

\bibitem{Dornheim_T_2022}
Dornheim T, B{\"o}hme M, Kraus D, D{\"o}ppner T, Preston T~R, Moldabekov Z~A and Vorberger J 2022 {\em Nat. Commun.\/} {\bf 13} 7911

\bibitem{Dornheim_T2_2022}
Dornheim T, Böhme M~P, Chapman D~A, Kraus D, Preston T~R, Moldabekov Z~A, Schlünzen N, Cangi A, Döppner T and Vorberger J 2023 {\em Phys. Plasmas\/} {\bf 30} 042707

\bibitem{dornheim2023xray}
Dornheim T, Döppner T, Baczewski A~D, Tolias P, Böhme M~P, Moldabekov Z~A, Ranjan D, Chapman D~A, MacDonald M~J, Preston T~R, Kraus D and Vorberger J 2023 {\em arXiv\/} (\textit{Preprint} \eprint{2305.15305})

\bibitem{Dornheim_insight_2022}
Dornheim T, Moldabekov Z, Tolias P, Böhme M and Vorberger J 2023 {\em Matter Radiat. Extrem.\/} {\bf 8} 056601

\bibitem{tolias2024fouriermatsubara}
Tolias P, Kalkavouras F and Dornheim T 2024 {\em J. Chem. Phys.\/} {\bf 160} 181102

\bibitem{kugler1}
Kugler A~A 1975 {\em J. Stat. Phys\/} {\bf 12} 35

\bibitem{Vitali_PRB_2010}
Vitali E, Rossi M, Reatto L and Galli D~E 2010 {\em Phys. Rev. B\/} {\bf 82}(17) 174510

\bibitem{Dornheim_PRB_nk_2021}
Dornheim T, B\"ohme M, Militzer B and Vorberger J 2021 {\em Phys. Rev. B\/} {\bf 103}(20) 205142

\bibitem{ISHTAR}
Dornheim T, Böhme M and Schwalbe S 2024 {ISHTAR - Imaginary-time Stochastic High- performance Tool for Ab initio Research} \urlprefix\url{https://doi.org/10.5281/zenodo.10497098}

\bibitem{repo}
 A link to a repository containing all PIMC raw data will be made available upon publication.

\bibitem{kugler_bounds}
Kugler A~A 1970 {\em Phys. Rev. A\/} {\bf 1} 1688

\bibitem{Baus_Hansen_OCP}
Baus M and Hansen J~P 1980 {\em Phys. Rep.\/} {\bf 59} 1--94

\bibitem{plasma2}
Ichimaru S 1982 {\em Rev. Mod. Phys\/} {\bf 54} 1017

\bibitem{Tolias_IEMHNC_2019}
Tolias P and Lucco~Castello F 2019 {\em Phys. Plasmas\/} {\bf 26} 043703

\bibitem{kugler_classical}
Kugler A~A 1973 {\em J. Stat. Phys.\/} {\bf 8} 107--153

\bibitem{Dornheim_PTR_2022}
Dornheim T, Vorberger J, Moldabekov Z~A and Böhme M 2023 {\em Phil. Trans. R. Soc. A\/} {\bf 381} 20220217

\bibitem{Dornheim_PRB_ESA_2021}
Dornheim T, Moldabekov Z~A and Tolias P 2021 {\em Phys. Rev. B\/} {\bf 103}(16) 165102

\bibitem{dornheim2024MatsubaraResponse}
Dornheim T, Tolias P, Kalkavouras F, Moldabekov Z and Vorberger J 2024 Dynamic exchange-correlation effects in the strongly coupled electron liquid (\textit{Preprint} \eprint{2405.08480})

\bibitem{Dornheim_PRR_2022}
Dornheim T, Vorberger J, Moldabekov Z~A and Tolias P 2022 {\em Phys. Rev. Research\/} {\bf 4}(3) 033018

\bibitem{Dornheim_JCP_ITCF_2021}
Dornheim T, Moldabekov Z~A and Vorberger J 2021 {\em J. Chem. Phys.\/} {\bf 155} 054110

\bibitem{Dornheim_JPSJ_2021}
Dornheim T, Vorberger J and Moldabekov Z~A 2021 {\em J. Phys. Soc. Jpn\/} {\bf 90} 104002

\end{thebibliography}

\end{document}